\documentclass[12pt]{article}
\usepackage{amssymb}
\usepackage{epsf}
\usepackage{epsfig}
\usepackage{cite}
\newcommand{\lsim}{
\mathrel{\hbox{\rlap{\hbox{\lower4pt\hbox{$\sim$}}}\hbox{$<$}}}}

\newcommand{\gsim}{
\mathrel{\hbox{\rlap{\hbox{\lower4pt\hbox{$\sim$}}}\hbox{$>$}}}}

\newcommand{\be}{\begin{equation}}
\newcommand{\ee}{\end{equation}}
\newcommand{\bi}{\begin{itemize}}
\newcommand{\ei}{\end{itemize}}
\newcommand{\ord}{{\cal O}}

\begin{document}
\begin{titlepage}
\vspace*{-0.5truecm}

\begin{flushright}
CERN-PH-TH/2005-241\\
TUM-HEP-610/05\\
MPP-2005-156\\
hep-ph/0512032
\end{flushright}

\vspace*{0.3truecm}

\begin{center}
\boldmath
{\Large{\bf New Aspects of $B\to\pi\pi,\pi K$\\
\vspace*{0.3truecm}
and their Implications for Rare Decays}}
\unboldmath
\end{center}

\vspace{0.9truecm}

\begin{center}
{\bf Andrzej J. Buras,${}^a$ Robert Fleischer,${}^b$ 
Stefan Recksiegel${}^a$ and Felix Schwab${}^{c,a}$}
 
\vspace{0.5truecm}

${}^a$ {\sl Physik Department, Technische Universit\"at M\"unchen,
D-85748 Garching, Germany}

\vspace{0.2truecm}

${}^b$ {\sl Theory Division, Department of Physics, CERN, 
CH-1211 Geneva 23, Switzerland}

\vspace{0.2truecm}

 ${}^c$ {\sl Max-Planck-Institut f{\"u}r Physik -- Werner-Heisenberg-Institut,
 D-80805 Munich, Germany}

\end{center}

\vspace{0.6cm}
\begin{abstract}
\vspace{0.2cm}\noindent
We analyse the $B\to\pi\pi,\pi K$ modes in the light of the most recent $B$-factory data, 
and obtain the following new results: (i) the $B^0_d\to\pi^+\pi^-,\pi^- K^+$ modes 
prefer $\gamma=(74\pm6)^\circ$, which -- together with $|V_{ub}/V_{cb}|$ -- allows
us to determine the ``true'' unitarity triangle and to search for CP-violating 
new-physics contributions to $B^0_d$--$\bar B^0_d$ mixing; (ii) the $B\to\pi K$ 
puzzle reflected in particular by the low experimental value of the ratio $R_{\rm n}$ 
of the neutral $B\to\pi K$ rates persists and still favours new physics in the electroweak
penguin sector with a new CP-violating phase $\phi \sim -90^\circ$,
although now also $\phi\sim +90^\circ$ can bring us rather close to the data; (iii) the 
mixing-induced $B^0_d\to\pi^0 K_{\rm S}$ CP asymmetry is a sensitive probe of the 
sign of this phase, and would currently favour $\phi\sim +90^\circ$, as well as the 
direct CP asymmetry of $B^\pm\to\pi^0K^\pm$, which suffers, however, from large 
hadronic uncertainties; (iv) we investigate the sensitivity of our $B\to\pi K$ analysis 
to large non-factorizable $SU(3)$-breaking effects and find that 
their impact is surprisingly small so that it is indeed exciting to speculate on 
new physics; (v) assuming that new physics enters through $Z^0$ penguins, we 
study the interplay between $B\to\pi K$ and rare $B$, $K$ decays and point out 
that the most recent $B$-factory constraints for the latter have interesting implications,
bringing us to a few scenarios for the future evolution of the data, where also the
mixing-induced CP violation in $B^0_d\to\pi^0 K_{\rm S}$ plays a prominent r\^ole.
\end{abstract}

\vspace*{0.5truecm}
\vfill
\noindent
December 2005

\end{titlepage}

\thispagestyle{empty}
\vbox{}
\newpage

\setcounter{page}{1}

\section{Introduction}\label{sec:intro}
Decays of $B$ mesons into $\pi\pi$ and $\pi K$ final states offer valuable 
information about strong interactions, weak interactions and possible new-physics 
(NP) effects. In a series of recent papers \cite{BFRS,BFRS-up}, we developed a 
strategy to address these aspects in a systematic manner. It uses the following 
working hypotheses:
\begin{itemize}
\item[(i)] $SU(3)$ flavour symmetry of strong interactions (but taking factorizable 
$SU(3)$-breaking corrections into account);
\item[(ii)] neglect of penguin annihilation and exchange topologies.
\end{itemize}
We may gain confidence in these assumptions through internal consistency checks, 
which worked well within the experimental uncertainties for our previous numerical analyses. Since the $B$ factories reported updated results for several of the input 
quantities, we would like to explore the implications for the 
picture emerging from our strategy. For a detailed overview of the current experimental
status of the $B\to\pi\pi,\pi K$ observables, we refer the reader to the most recent
compilation of the Heavy Flavour Averaging Group (HFAG) \cite{HFAG}. We will 
give the updated numerical values for the quantities entering our strategy below.

A somewhat surprising new development of this summer is a new world average 
for $(\sin 2\beta)_{\psi K_{\rm {S}}}$, which went down by about 1$\sigma$ thanks 
to an update by the Belle collaboration \cite{new-Belle-s2b}, and is now given as follows:
\begin{equation}\label{s2b-exp}
(\sin 2\beta)_{\psi K_{\rm {S}}}=0.687\pm0.032, \qquad
\beta=(21.7^{+1.3}_{-1.2})^\circ.
\end{equation}
The corresponding straight line in the $\bar\rho$--$\bar\eta$ plane of the generalized
Wolfenstein parameters \cite{wolf,blo} is now on the lower side 
of the allowed region for the apex of the unitarity triangle (UT) of the 
Cabibbo--Kobayashi--Maskawa (CKM) matrix that follows from the usual ``indirect" 
fits \cite{UTfit,CKMfitter}. In view of this result, we assume that this potential
discrepancy is due to NP in $B^0_d$--$\bar B^0_d$ mixing, and perform an
analysis of the UT in Section~\ref{sec:gam-UT}. To this end, we use the data for
the decays $B^0_d\to\pi^+\pi^-$ and $B^0_d\to\pi^-K^+$, which allow us to 
determine the ``true" UT angle $\gamma$ \cite{gamma-det,FleischerMatias},  
serving as an input for our subsequent analysis. Complementing this information 
with the measurement of $|V_{ub}/V_{cb}|$ through semi-leptonic $B$ decays, 
we can construct the so-called reference unitarity triangle \cite{GNW,refut}, and are 
in a position to convert the possibly emerging discrepancy for the UT into a CP-violating 
NP physics phase in $B^0_d$--$\bar B^0_d$ mixing. Moreover, we may  extract
the ``true" values of $\alpha$ and $\beta$, where the latter serves as
an input for our analysis of rare decays. 

In Section~\ref{sec:Bpipi}, we then extract the hadronic parameters 
characterizing the $B\to\pi\pi$ system with the help of the $SU(2)$ isospin
symmetry of strong interactions, and predict the CP-violating observables of 
the $B^0_d\to\pi^0\pi^0$ channel. The results of this section are essentially
theoretically clean, and serve as a testing ground for the calculation of
the dynamics of the $B\to\pi\pi$ decays directly from QCD-related approaches,
such as ``QCD factorization'' (QCDF) \cite{QCDF}, the perturbative QCD 
approach (PQCD) \cite{PQCD}, ``soft collinear effective theory" (SCET) \cite{SCET}, 
or QCD sum rules \cite{sum-rules}.

The $B\to\pi K$ system and the status of the
``$B\to\pi K$ puzzle" are then the subject of Section~\ref{sec:BpiK}. We
find that our Standard-Model (SM) predictions for those decays that are only
marginally affected by electroweak (EW) penguins are in accordance with the experimental picture, whereas this is not the case for the observables with prominent contributions from these topologies, in particular for the ratio $R_{\rm n}$ 
of the CP-averaged rates of the neutral $B\to\pi K$ modes. We show that 
this puzzle can still be resolved through
NP effects in the EW penguin sector with a large CP-violating new phase $\phi$. 
We have also a closer look at another hot topic -- the mixing-induced CP 
asymmetry of the $B^0_d\to \pi^0K_{\rm S}$ channel -- and point out that this 
quantity depends strongly on the sign of the NP phase $\phi$. In particular,
this asymmetry, which is found experimentally to be significantly smaller than the 
SM expectation, can be brought closer to the data by reversing the sign of 
$\phi$. Moreover, we investigate whether significant non-factorizable 
$SU(3)$-breaking effects could have large impact on our analysis. Interestingly, 
we find that this is not the case, and note that such effects are also not indicted 
by the internal consistency checks of our working assumptions. 

In Section~\ref{sec:rare-BK-scenarios}, we explore the interplay of the
NP in the EW penguin contributions to the $B\to\pi K$ system with rare $B$ and 
$K$ decays. To this end, we apply the popular scenario that NP enters the
EW penguins through $Z^0$-penguin topologies
\cite{Colangelo:1998pm,Buras:1998ed,BRS,Buras:1999da,Buchalla:2000sk}.
In view of new experimental results, we speculate on possible future scenarios. 
As in our previous analysis, we find that $K\to\pi\nu\bar\nu$, 
$K_{\rm L}\to\pi^0e^+e^-$ and
$B_{s,d}\to\mu^+\mu^-$ are sensitive probes for these scenarios. This
is also the case for the mixing-induced  $B^0_d\to \pi^0K_{\rm S}$ CP asymmetry
discussed in Section~\ref{sec:BpiK}. Finally, we summarize our
conclusions in Section~\ref{sec:concl}.

\boldmath
\section{Analysis of the Unitarity Triangle}\label{sec:gam-UT}
\unboldmath
Let us in view of the new result for $(\sin 2\beta)_{\psi K_{\rm {S}}}$ first have a 
closer look at the UT. The starting point of this analysis is the assumption that 
the possible discrepancy between (\ref{s2b-exp}) and the CKM fits is due to 
CP-violating NP contributions to $B^0_d$--$\bar B^0_d$ mixing (although it is 
of course too early to say something definite on this issue at the moment) 
\cite{GNW,FM-BpsiK,FIM}. For other recent analyses in this context, see
\cite{UTfit,CKMfitter,Agashe:2005hk}.
We may then extract the general $B^0_d$--$\bar B^0_d$ mixing phase
\begin{equation}\label{phid-def}
\phi_d=\phi_d^{\rm SM}+\phi_d^{\rm NP}=2\beta+\phi_d^{\rm NP},
\end{equation}
where $\phi_d^{\rm NP}$ could originate from physics beyond the SM, from the
numerical value in (\ref{s2b-exp}), yielding $\phi_d=(43.4^{+2.6}_{-2.4})^\circ$.
Here we have discarded a possible second solution around $136.6^\circ$ 
\cite{FleischerMatias,FIM},
which is disfavoured by recent $B$-factory data \cite{HFAG}. The phase
(\ref{phid-def}) enters the mixing-induced CP asymmetry of the 
$B_d\to\pi^+\pi^-$ channel, which arises in the following time-dependent
rate asymmetry:
\begin{eqnarray}
\lefteqn{\frac{\Gamma(B^0_d(t)\to \pi^+\pi^-)-\Gamma(\bar B^0_d(t)\to 
\pi^+\pi^-)}{\Gamma(B^0_d(t)\to \pi^+\pi^-)+\Gamma(\bar B^0_d(t)\to 
\pi^+\pi^-)}}\nonumber\\
&&={\cal A}_{\rm CP}^{\rm dir}(B_d\to \pi^+\pi^-)\cos(\Delta M_d t)+
{\cal A}_{\rm CP}^{\rm mix}(B_d\to \pi^+\pi^-)
\sin(\Delta M_d t).\label{rate-asym}
\end{eqnarray}
In the SM, these observables can be written as 
(for the explicit expressions, see \cite{gamma-det})
\begin{eqnarray}
{\cal A}_{\rm CP}^{\rm dir}(B_d\to \pi^+\pi^-)&=& G_1(d,\theta;\gamma) 
 \,\stackrel{{\rm exp}}{=}\, -0.37\pm0.10\label{CP-Bpipi-dir-gen}\\
{\cal A}_{\rm CP}^{\rm mix}(B_d\to \pi^+\pi^-)&=& G_2(d,\theta;\gamma,\phi_d)
 \,\stackrel{{\rm exp}}{=}\, +0.50\pm0.12,\label{CP-Bpipi-mix-gen}
\end{eqnarray}
where we have also given the most recent experimental numbers, and 
$de^{i\theta}$ is a CP-conserving hadronic parameter, which measures --
sloppily speaking -- the ratio of the $B_d\to \pi^+\pi^-$ penguin to tree contributions.

Let us now use the additional information which is provided by the
$B_d\to\pi^\mp K^\pm$ decays. The assumptions listed at the beginning of
Section~\ref{sec:intro}  allow us then to derive 
\begin{equation}\label{H-rel}
H_{\rm BR}\equiv\underbrace{\frac{1}{\epsilon}
\left(\frac{f_K}{f_\pi}\right)^2\left[\frac{\mbox{BR}
(B_d\to\pi^+\pi^-)}{\mbox{BR}(B_d\to\pi^\mp K^\pm)}
\right]}_{\mbox{$7.5\pm 0.7$}} =
\underbrace{-\frac{1}{\epsilon}\left[\frac{{\cal A}_{\rm CP}^{\rm dir}(B_d\to\pi^\mp 
K^\pm)}{{\cal A}_{\rm CP}^{\rm dir}(B_d\to\pi^+\pi^-)}
\right]}_{\mbox{$6.7\pm 2.0$}} \equiv H_{{\cal A}_{\rm CP}^{\rm dir}},
\end{equation}
where $\epsilon\equiv\lambda^2/(1-\lambda^2)=0.053$, and the ratio
$f_K/f_\pi=160/131$ of the kaon and pion decay constants takes factorizable
$SU(3)$-breaking corrections into account. In (\ref{H-rel}), we have indicated the numerical values following from the current data. Consequently, within the 
experimental uncertainties, this relation is also well satisfied by the new data, 
which gives us further confidence in our working assumptions. 

\begin{figure}
\begin{center}
\includegraphics[width=8cm]{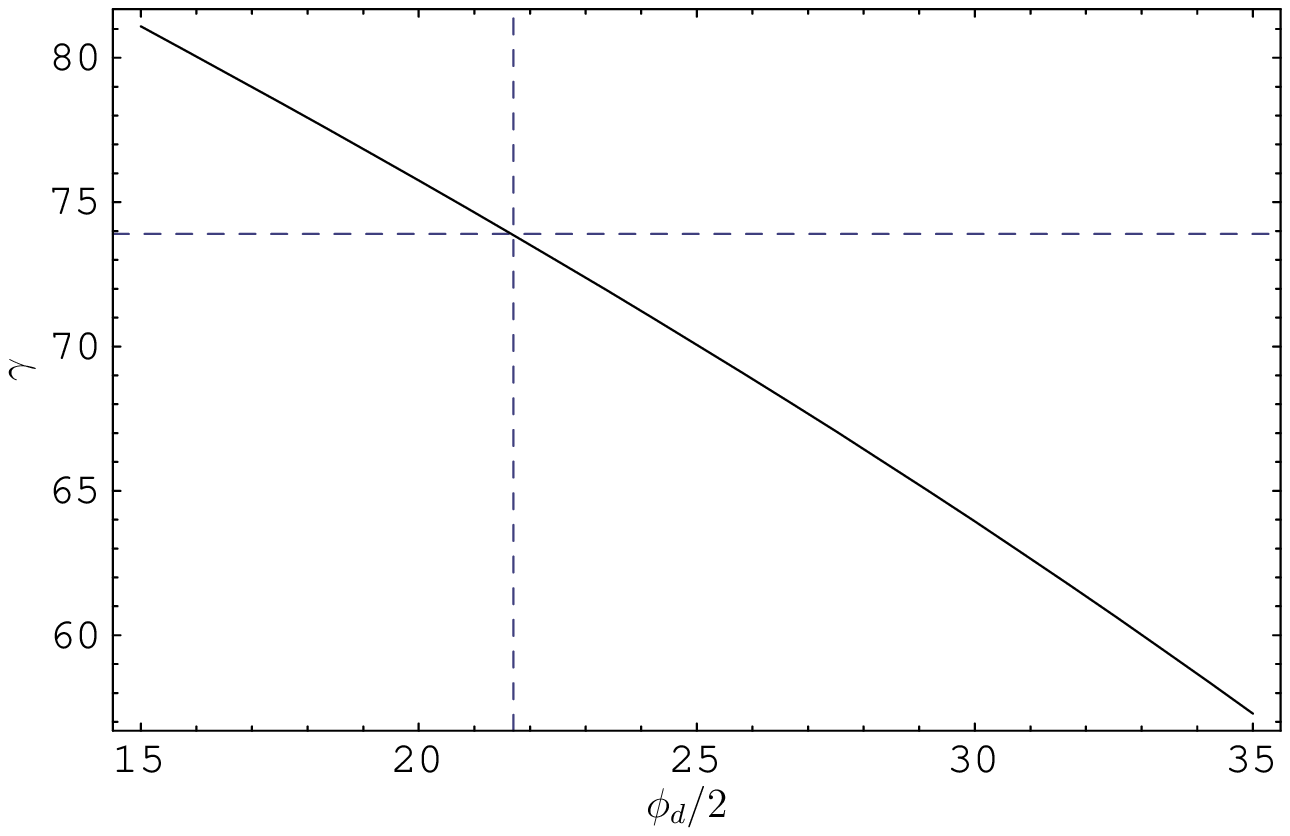}
\end{center}
\vspace*{-0.5truecm}
\caption{The value of $\gamma$ as determined in our strategy as a function
of $\phi_d/2$ with all other experimental input parameters kept at
their central values. The lines correspond to the values in (\ref{s2b-exp}) and
(\ref{gamma-det}).\label{fig:beta-gamma}}
\end{figure}

The quantities $H_{\rm BR}$ and $H_{{\cal A}_{\rm CP}^{\rm dir}}$, which are 
fixed through the branching ratios and direct CP asymmetries, respectively, 
can be written as follows:
\begin{equation}\label{H-fct}
H_{\rm BR} = G_3(d,\theta;\gamma) =
H_{{\cal A}_{\rm CP}^{\rm dir}}.
\end{equation}
If we complement this expression with (\ref{CP-Bpipi-dir-gen}) and
(\ref{CP-Bpipi-mix-gen}), we have sufficient
information to determine $\gamma$, as well as $(d,\theta)$  
\cite{gamma-det,FleischerMatias}. Following these lines yields
\begin{equation}\label{BR-det}
\left.\gamma\right|_{\rm BR}=(44.0^{+4.3}_{-3.7})^\circ \quad\lor\quad  
(70.1^{+5.6}_{-7.2})^\circ,
\end{equation}
\begin{equation}\label{ACP-det}
\left.\gamma\right|_{{\cal A}_{\rm CP}^{\rm dir}}=(42.1^{+3.4}_{-3.6})^\circ 
\quad\lor\quad  (73.9^{+5.8}_{-6.5})^\circ.
\end{equation}
Consequently, $H_{\rm BR}$ and $H_{{\cal A}_{\rm CP}^{\rm dir}}$ give results 
that are in good agreement with one another. As we discussed in \cite{BFRS},
the solutions around $40^\circ$ can be excluded through an analysis of
the whole $B\to\pi\pi,\pi K$ system, which is also the case for the most 
recent data. In the following analysis, we will use 
\begin{equation}\label{gamma-det}
\gamma=(73.9^{+5.8}_{-6.5})^\circ,
\end{equation}
corresponding to $H_{{\cal A}_{\rm CP}^{\rm dir}}$, as this is theoretically cleaner than
the avenue offered by $H_{\rm BR}$. As we will see in Subsection~\ref{ssec:SU3break}, 
even large non-factorizable $SU(3)$-breaking corrections have a remarkably small
impact on the numerical result in (\ref{gamma-det}). 
The value for $\gamma$ in (\ref{gamma-det}) is somewhat larger than
in \cite{BFRS}, a significant part of the numerical shift can be
explained by the new value for $(\sin 2\beta)_{\psi K_{\rm {S}}}$,
as shown in Fig.~\ref{fig:beta-gamma}.

Before having a closer look at the whole set of hadronic parameters
characterizing the $B\to\pi\pi$ system, let us first explore the implications 
of (\ref{gamma-det}) for the apex of the UT in the 
$\bar\rho$--$\bar\eta$ plane. The interesting feature of this value of $\gamma$
following from the CP asymmetries of the $B_d\to \pi^+\pi^-$, $B_d\to\pi^\mp K^\pm$
system is that it does not receive -- in our scenario of NP -- any significant NP 
contributions. Consequently, it is the ``true" angle $\gamma$ of the UT. 
In order to complete the determination of the ``true'' UT, i.e.\ of the so-called 
reference UT \cite{GNW,refut}, we use the ratio $|V_{ub}/V_{cb}|$ extracted from 
semi-leptonic tree-level $B$ decays. Although its values extracted from exclusive 
and inclusive decays are markedly different from each other, we use the 
following average \cite{UTfit}:
\begin{equation}\label{vub}
\left|\frac{V_{ub}}{V_{cb}}\right|= 0.102\pm  0.005.
\end{equation}
In the second column of  Table~\ref{tab:RUT}, we list the values of $\bar\rho$,
 $\bar\eta$, and ``$\beta_{\rm true}$" for this value of $|V_{ub}/V_{cb}|$ and 
$\gamma$ in (\ref{gamma-det}). For completeness, we also give the values of
the lengths of the UT sides $R_b$ and $R_t$ and of the angle $\alpha$. 
We observe that the value 
of $\beta_{\rm true}$ that we obtain this way is significantly
higher than the one in (\ref{s2b-exp}). It corresponds to 
$(\sin 2\beta)_{\rm true}=0.78\pm 0.03$.

In our scenario, this difference is attributed to a non-vanishing value of
the NP phase $\phi_d^{\rm NP}$ in (\ref{phid-def}), where $\beta$ corresponds
to $\beta_{\rm true}$. As seen in Table~\ref{tab:RUT},  our value of 
$\phi_d^{\rm NP}$ is 
compatible with the one found in \cite{UTfit,CKMfitter}, but our value of $\gamma$ 
obtained here in a different manner is significantly higher. 
An even larger value of $\gamma$ following from $B \to \pi \pi$, in the ballpark of $80^{\circ}$,
has been reported from an analysis using SCET \cite{SCET-Bdpi0K0}.
In Table~\ref{tab:RUT}, we also show the results for
the reference unitarity triangle (RUT) obtained in \cite{UTfit}, 
where only CP violation in
$B\to DK$ and $|V_{ub}/V_{cb}|$ in (\ref{vub}) have been used
as input. The agreement between our analysis and the one in \cite{UTfit}
is almost too good.
In obtaining the values in the column ``UTfit RUT'' we have used
$\bar\rho$ and $\bar\eta$ from \cite{UTfit}.
In the last two columns in Table \ref{tab:RUT} we
collect the results from \cite{UTfit} within the SM and for the universal unitarity
triangle (UUT) in models with minimal flavour violation
\cite{Buras:2000dm,D'Ambrosio:2002ex}, where (\ref{s2b-exp}), 
$|V_{ub}/V_{cb}|$ and $\Delta M_d/\Delta M_s$ serve as inputs. As
already stated above, the values of $\gamma$ are in both cases significantly smaller, while
the values of $\alpha$ are significantly larger than in the case of the RUT.

The visibly increased value of $R_t$ relatively to the standard UT fits found
by us in the case of the RUT
would require a small {\it negative} NP contribution to the 
$B^0_d$--$\bar B^0_d$ mass difference $\Delta M_d$ and/or a slightly increased
value of the non-perturbative parameter $\xi$ relevant for the ratio 
$\Delta M_d/\Delta M_s$. We look forward to improved data on the $B \to \pi \pi$,
$B \to \pi K$ system, $|V_{ub}/V_{cb}|$, $\sin 2 \beta$ and $\Delta M_s$ in order to see whether the difference between
the large value of $\gamma$ found here and the one resulting from the UUT and full UT fits could be interpreted as a clear signal of NP. 

\begin{table}
\vspace{0.4cm}
\begin{center}
\begin{tabular}{|c||c|c|c|c|c|}
\hline
  Quantity &  Our Value & UTfit RUT & Full UT & UUT \\ \hline
$\gamma$ & $(73.9^{+5.8}_{-6.5})^\circ$ & $(65\pm18)^\circ$ & $(57.6 \pm 5.5)^\circ$  & $(51 \pm 10)^\circ$\\\hline
$\bar\rho$ & $0.127 \pm 0.046$ & $0.18\pm0.12$& $0.216\pm0.036$  & $0.259 \pm 0.068$  \\\hline
$\bar\eta$ & $0.422 \pm 0.025$ & $0.41\pm0.05$& $0.342 \pm 0.022$ & $0.320 \pm 0.042$ \\\hline
$R_b$ & $0.44 \pm 0.02$ & $0.45 \pm 0.07$ & $0.40 \pm 0.03  $   &  $0.42 \pm 0.05 $\\\hline
$R_t$ & $0.97 \pm 0.05$ & $0.92 \pm 0.11$ & $0.86\pm 0.03$  & $0.81 \pm 0.06$ \\\hline
$\beta_{\rm true}$ & $(25.8\pm 1.3)^\circ$  & $(26.1\pm 3.0)^\circ$ & $23.8 \pm 1.5 $  & $23.4 \pm 1.3$\\\hline
$\alpha$ & $( 80.3 ^{+6.6}_{-5.9} )^\circ$  & $( 87 \pm 15 )^\circ$ & $( 98.5 \pm 5.7)^\circ $  & $( 105 \pm 11 )^\circ$\\\hline
$(\sin 2\beta)_{\rm true}$ & $0.782\pm0.029$ & $0.782\pm0.065$ &  $0.735 \pm 0.024$ & $0.728 \pm 0.031$ 
\\\hline
$\phi_d^{\rm NP}$  & $-(8.2\pm 3.5)^\circ $ &  $-(8.9\pm 6.0)^\circ $ &  $-(4.1 \pm3.9)^\circ $  & $-(3.3 \pm 3.6)^\circ$ \\\hline
\end{tabular}
\end{center}
\caption{Parameters of the reference UT (RUT) determined through 
$|V_{ub}/V_{cb}|$ in (\ref{vub}) and the CP asymmetries 
of the $B_d\to \pi^+\pi^-$, $B_d\to\pi^\mp K^\pm$ system, yielding the value of 
$\gamma$ in (\ref{gamma-det}), compared with the results of \cite{UTfit}. We show also the results
of the full UT fit and of the universal unitarity triangle obtained in \cite{UTfit}.
}\label{tab:RUT}
\end{table}

\boldmath
\section{The $B\to\pi\pi$ System}\label{sec:Bpipi}
\unboldmath
Let us now continue the analysis of the $B\to\pi\pi$ system. In addition to
the CP-violating observables in (\ref{rate-asym}), we use the following 
ratios of CP-averaged branching ratios:
\begin{eqnarray}
R_{+-}^{\pi\pi}&\equiv&2\left[\frac{\mbox{BR}(B^+\to\pi^+\pi^0)
+\mbox{BR}(B^-\to\pi^-\pi^0)}{\mbox{BR}(B_d^0\to\pi^+\pi^-)
+\mbox{BR}(\bar B_d^0\to\pi^+\pi^-)}\right]\stackrel{{\rm exp}}{=}2.04\pm0.28
\label{Rpm-def}\\
R_{00}^{\pi\pi}&\equiv&2\left[\frac{\mbox{BR}(B_d^0\to\pi^0\pi^0)+
\mbox{BR}(\bar B_d^0\to\pi^0\pi^0)}{\mbox{BR}(B_d^0\to\pi^+\pi^-)+
\mbox{BR}(\bar B_d^0\to\pi^+\pi^-)}\right]\stackrel{{\rm exp}}{=}0.58\pm0.13.
\end{eqnarray}
Using the isospin symmetry of strong interactions, these quantities can be 
written as
\begin{equation}\label{Rpipi-gen}
R_{+-}^{\pi\pi}=F_1(d,\theta,x,\Delta;\gamma), \quad
R_{00}^{\pi\pi}=F_2(d,\theta,x,\Delta;\gamma),
\end{equation}
where $xe^{i\Delta}$ is another hadronic parameter, which was introduced
in \cite{BFRS}. Using now, in addition, the CP-violating observables in
(\ref{CP-Bpipi-dir-gen}) and (\ref{CP-Bpipi-mix-gen}) and the value of 
$\gamma$ in (\ref{gamma-det}), 
we arrive at the following set of hadronic parameters:
\begin{equation}\label{d-theta-det}
d=0.52^{+0.09}_{-0.09}\quad[0.51^{+0.26}_{-0.20}], \quad 
\theta=(146^{+7.0}_{-7.2})^\circ\quad[(140^{+14}_{-18})^\circ]
\end{equation}
\begin{equation}\label{x-Delta-det}
x=0.96^{+0.13}_{-0.14}\quad[1.15^{+0.18}_{-0.16}], \quad 
\Delta=-(53^{+18}_{-26})^\circ\quad[-(59^{+19}_{-26})^\circ],
\end{equation}
which is in excellent agreement with the picture of our last analysis in
\cite{BFRS-up}, corresponding to the numbers in parentheses. As in this
paper, we include also the EW penguin effects in the $B\to\pi\pi$ system
\cite{BF98,GPY}, although these topologies have a tiny impact \cite{PAPIII}.
Let us emphasize that the results for the hadronic parameters listed
above, which are consistent with the analyses of
other authors (see, for instance, \cite{ALP-Bpipi,CGRS,WuZhou}), 
are essentially theoretically clean and serve as a testing ground for 
calculations within QCD-related approaches, such as QCDF \cite{QCDF}, 
PQCD \cite{PQCD}, SCET \cite{SCET},  or QCD sum rules \cite{sum-rules}.

Finally, we can predict the CP asymmetries of the decay 
$B_d\to\pi^0\pi^0$, where we obtain 
\begin{equation}\label{ACP-Bdpi0pi0-pred}
{\cal A}_{\rm CP}^{\rm dir}(B_d\to \pi^0\pi^0)=-0.30^{+0.48}_{-0.26}, \quad
{\cal A}_{\rm CP}^{\rm mix}(B_d\to \pi^0\pi^0)=-0.87^{+0.29}_{-0.19}.
\end{equation}
On the other hand, the current experimental value for the direct CP 
asymmetry is \cite{HFAG}:
\begin{equation}\label{ACP-Bdpi0pi0-exp}
{\cal A}_{\rm CP}^{\rm dir}(B_d\to \pi^0\pi^0)=-0.28^{+0.40}_{-0.39}.
\end{equation}
No stringent test of our predictions is
provided at this stage, but the indicated agreement is very encouraging.

\boldmath
\section{The $B\to\pi K$ System}\label{sec:BpiK}
\unboldmath
Following our strategy developed in \cite{BFRS}, we are now in a position to 
calculate the observables of the $B\to\pi K$ system in the SM.
The corresponding decays fall into two classes: transitions with a negligible
impact of EW penguins, and channels receiving sizeable contributions from
these topologies.

\boldmath
\subsection{The Decays $B_d\to\pi^\mp K^\pm$ and $B^\pm\to\pi^\pm K$}
\unboldmath
Let us first have a look at those decays that are marginally affected by
contributions from EW penguin diagrams, $B_d\to\pi^\mp K^\pm$
and $B^\pm\to\pi^\pm K$. We encountered the former channel already in
the SM relation (\ref{H-rel}), which is satisfied by the current 
data. Concerning the latter decay, it provides the CP-violating asymmetry
\begin{equation}\label{ACP-BppipK0}
{\cal A}_{\rm CP}^{\rm dir}(B^\pm\to\pi^\pm K)\equiv
\frac{\mbox{BR}(B^+\to\pi^+K^0)-
\mbox{BR}(B^-\to\pi^-\bar K^0)}{\mbox{BR}(B^+\to\pi^+K^0)+
\mbox{BR}(B^-\to\pi^-\bar K^0)} \,\stackrel{{\rm exp}}{=}\, 0.02 \pm 0.04,
\end{equation}
and enters in the following ratio \cite{FM}:
\begin{equation}\label{R-def}
R\equiv\left[\frac{\mbox{BR}(B_d^0\to\pi^- K^+)+
\mbox{BR}(\bar B_d^0\to\pi^+ K^-)}{\mbox{BR}(B^+\to\pi^+ K^0)+
\mbox{BR}(B^-\to\pi^- \bar K^0)}
\right]\frac{\tau_{B^+}}{\tau_{B^0_d}} \,\stackrel{{\rm exp}}{=}\, 0.86\pm0.06;
\end{equation}
the numerical values refer again to the most recent compilation of the
HFAG in \cite{HFAG}. 
The $B^+\to\pi^+ K^0$ channel involves another hadronic parameter,
$\rho_{\rm c}e^{i\theta_{\rm c}}$, which cannot be determined through
the $B\to\pi\pi$ data \cite{BF98,defan,neubert}:
\begin{equation}\label{B+pi+K0}
A(B^+\to\pi^+K^0)=-P'\left[1+\rho_{\rm c}e^{i\theta_{\rm c}}e^{i\gamma}
\right];
\end{equation}
the overall normalization $P'$ cancels in (\ref{ACP-BppipK0}) in
(\ref{R-def}). Usually, it is  
assumed that $\rho_{\rm c}e^{i\theta_{\rm c}}$ can be neglected.
In this case, the direct
CP asymmetry in (\ref{ACP-BppipK0}) vanishes, and $R$ can be calculated
through the $B\to\pi\pi$ data with the help of the assumptions specified
at the beginning of Section~\ref{sec:intro}:
\begin{equation}\label{R-pred-0}
R|_{\rm SM}=0.963^{+0.019}_{-0.022}.
\end{equation}

This numerical result is $1.6 \sigma$ larger than the experimental value
in (\ref{R-def}). As we discussed in detail in \cite{BFRS-up}, 
the experimental range for the direct CP asymmetry in (\ref{ACP-BppipK0})  
and the first direct signals for $B^\pm\to K^\pm K$ decays \cite{FR} favour a 
value of $\theta_{\rm c}$ around $0^\circ$. This feature allows us to essentially 
resolve the small discrepancy concerning $R$ for values of $\rho_{\rm c}$ around 
0.05. The remaining small numerical difference between the calculated value of
$R$ and the experimental result, if confirmed by future data, could be due to
(small) colour-suppressed EW penguins, which enter $R$ as well \cite{BFRS}.
As we will see in Subsection~\ref{ssec:SU3break}, even large non-factorizable
$SU(3)$-breaking effects would have a small impact on the predicted value 
of $R$. In view of these results, we would not be surprised to see an increase 
of the experimental value of $R$ in the future.

\boldmath
\subsection{The Decays $B^\pm\to\pi^0 K^\pm$ and 
$B_d\to\pi^0 K$}\label{ssec:BpiK-EWP}
\unboldmath
Let us now turn to those $B\to\pi K$ modes that are significantly 
affected by EW penguin contributions, the $B^+\to\pi^0K^+$ and
$B^0_d\to\pi^0K^0$ channels. The key observables for the exploration
of these modes are the following ratios of their CP-averaged branching
ratios \cite{BF98,BF00}:
\begin{equation}\label{Rc-def}
R_{\rm c}\equiv2\left[\frac{\mbox{BR}(B^+\to\pi^0K^+)+
\mbox{BR}(B^-\to\pi^0K^-)}{\mbox{BR}(B^+\to\pi^+ K^0)+
\mbox{BR}(B^-\to\pi^- \bar K^0)}\right] \,\stackrel{{\rm exp}}{=}\, 1.01\pm0.09
\end{equation}
\begin{equation}\label{Rn-def}
R_{\rm n}\equiv\frac{1}{2}\left[
\frac{\mbox{BR}(B_d^0\to\pi^- K^+)+
\mbox{BR}(\bar B_d^0\to\pi^+ K^-)}{\mbox{BR}(B_d^0\to\pi^0K^0)+
\mbox{BR}(\bar B_d^0\to\pi^0\bar K^0)}\right] \,\stackrel{{\rm exp}}{=}\, 0.83\pm0.08.
\end{equation}
The EW penguin effects are described by a parameter $q$, which measures
the strength of the EW penguins with respect to tree-diagram-like topologies,
and a CP-violating phase $\phi$. In the SM, this phase vanishes,
and $q$ can be calculated with the help of the $SU(3)$ flavour symmetry,
yielding a value of $0.69 \cdot 0.086/|V_{ub}/V_{cb}|= 0.58$ \cite{NR}.
We find then
\begin{equation}\label{RncSM}
R_{\rm c}|_{\rm SM}=1.15 \pm 0.05, \quad R_{\rm n}|_{\rm SM}=1.12 \pm 0.05.
\end{equation}

\begin{figure}
\begin{center}
\includegraphics[width=12cm]{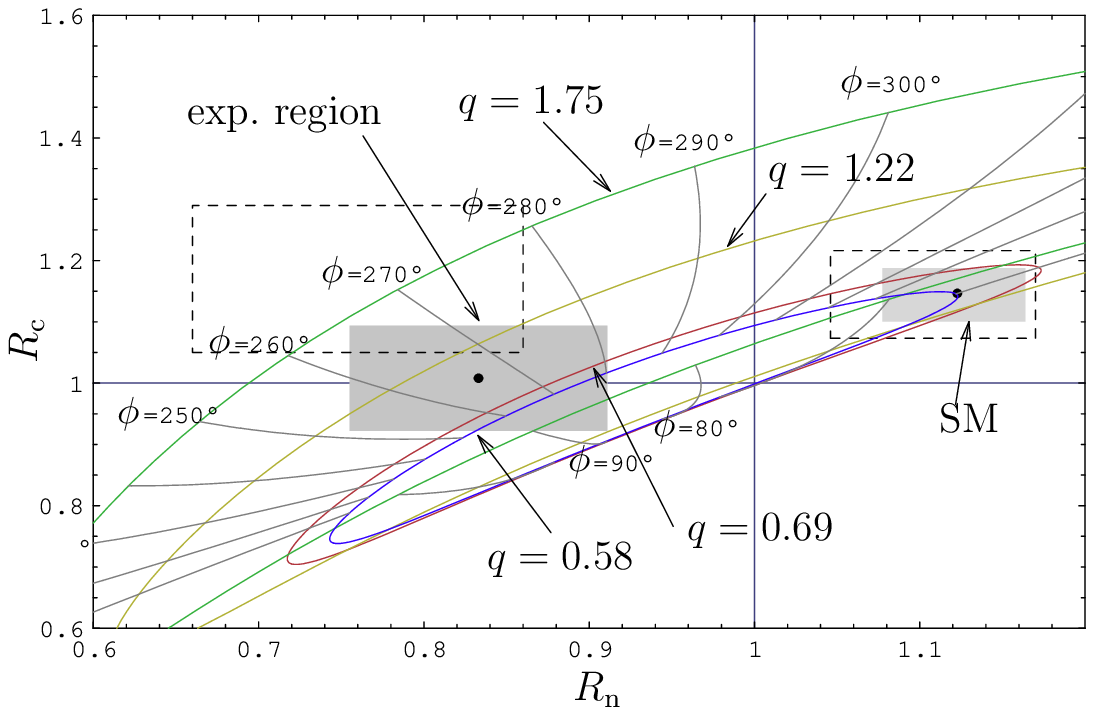}
\end{center}
\vspace*{-0.5truecm}
\caption{The current situation in the $R_{\rm n}$--$R_{\rm c}$ plane: the shaded 
areas indicate the experimental and SM $1 \sigma$ ranges, the lines show the
theory predictions for the central values of the hadronic parameters
and various values of $q$ with $\phi\in[0^\circ,360^\circ]$.
The plot ranges and the displayed values of $q$ correspond to those considered in
\cite{BFRS}.}
\label{fig:RnRc}
\end{figure}

Following \cite{BFRS}, we discuss the dependence of $R_{\rm n}$ and 
$R_{\rm c}$ on $q$ and $\phi$ with the help of a plot of the $R_{\rm n}$--$R_{\rm c}$ 
plane (Fig.~\ref{fig:RnRc}). The experimental range is still far from the
SM predictions; for the convenience of the reader we have indicated
the experimental range and the SM predictions at the time of our
original analysis \cite{BFRS} with dashed rectangles. Although
the central values of $R_{\rm n}$ and $R_{\rm c}$ have slightly moved towards
each other, the puzzle is as prominent as ever. The experimental
region can now be reached without an enhancement of $q$, but
a large CP-violating phase $\phi$ of the order of $-90^\circ$ is
still required, although $\phi$ of the order of $+90^\circ$ can
also bring us rather close to the experimental range of $R_{\rm n}$ and 
$R_{\rm c}$. We will return to this alternative below. Explicitly, we find
\begin{equation}
\label{q-phi}
q=0.99\,^{+0.66}_{-0.70} ,\quad \phi=-(94\,^{+16}_{-17} )^\circ.
\end{equation}
The impact of rare decays on these values will be discussed in 
Section~\ref{sec:rare-BK-scenarios}, where various scenarios with different 
values of $q$ and $\phi$ will be considered.

\boldmath
\subsection{CP Violation in  $B_d\to\pi^0 K_{\rm S}$ and 
$B^\pm\to\pi^0K^\pm$}\label{ssec:BpiK-CP}
\unboldmath
In the SM, the CP asymmetries of the decay $B_d\to\pi^0 K_{\rm S}$, which
can be extracted from a time-dependent rate asymmetry of the same
form as (\ref{rate-asym}), are expected to satisfy the following relations \cite{PAPIII}:
\begin{equation}\label{Bdpi0K0-rel}
{\cal A}_{\rm CP}^{\rm dir}(B_d\!\to\!\pi^0 K_{\rm S})\approx 0, \quad
\underbrace{{\cal A}_{\rm CP}^{\rm mix}(B_d\!\to\!\pi^0 K_{\rm S})}_{\mbox{$\equiv
-(\sin2\beta)_{\pi^0K_{\rm S}}$}}\approx
\underbrace{{\cal A}_{\rm CP}^{\rm mix}(B_d\!\to\!\psi K_{\rm S})}_{\mbox{$\equiv
-(\sin2\beta)_{\psi K_{\rm S}}$}}.
\end{equation}
The most recent $B$-factory results read as follows \cite{HFAG}:
\begin{equation}\label{Bdpi0K0-exp}
{\cal A}_{\rm CP}^{\rm dir}(B_d\!\to\!\pi^0 K_{\rm S})=-0.02\pm0.13, \quad
{\cal A}_{\rm CP}^{\rm mix}(B_d\!\to\!\pi^0 K_{\rm S})=-0.31\pm0.26,
\end{equation}
where the BaBar and Belle collaborations are in agreement with each other. 
Comparing with (\ref{s2b-exp}), we see that there is a sizeable departure of the
experimentally measured value of $(\sin2\beta)_{\pi^0K_{\rm S}}$ from
$(\sin2\beta)_{\psi K_{\rm S}}$, which is one of the recent hot topics. 

Consequently,
a detailed theoretical analysis of the relations in (\ref{Bdpi0K0-rel}) is required. In fact, 
our strategy developed in \cite{BFRS} allows us to address this issue and to
{\it predict} the CP-violating observables of the $B_d\to\pi^0 K_{\rm S}$ channel both 
within the SM and within our scenario of NP discussed above. A detailed analysis along 
these lines was already presented by us in \cite{BFRS}, from which one can extract
\begin{equation}\label{DS}
\Delta S \equiv (\sin2\beta)_{\pi^0K_{\rm S}}-
(\sin2\beta)_{\psi K_{\rm S}} \,\stackrel{\rm exp}{=}\, -0.38\pm 0.26
\end{equation}
to be {\it positive} in the SM, and in the ballpark of $0.10$--$0.15$. The
difference introduced in (\ref{DS}) allows a direct comparison with the results
obtained in the literature, where values for $\Delta S$ in the range
$0.04-0.08$ can be found that were obtained within the context of 
dynamical approaches like QCDF \cite{beneke-Bdpi0K0} 
and SCET \cite{SCET-Bdpi0K0}. Moreover, bounds were derived with
the help of the $SU(3)$ flavour symmetry \cite{SU3-bounds}. 
Using the formulae of Section 4.5 in \cite{BFRS}, our updated values for 
the CP asymmetries in  $B_d\to\pi^0 K_{\rm S}$ within the SM read as follows:
\begin{equation}\label{CPpi0KS}
{\cal A}_{\rm CP}^{\rm dir}(B_d\!\to\!\pi^0 K_{\rm S})|_{\rm SM}=0.06^{+0.09}_{-0.10},
\qquad
{\cal A}_{\rm CP}^{\rm mix}(B_d\!\to\!\pi^0 K_{\rm S})|_{\rm
  SM}=-(0.82^{+0.03}_{-0.04}).
\end{equation}
Consequently, we find
\begin{equation}\label{DSSM}
\Delta S\vert_{\rm SM}= 0.13\pm0.05,
\end{equation}
in agreement with other estimates but somewhat larger. We stress that in
obtaining this result we did {\it not} have to rely on dynamical frameworks 
that use ideas of factorization, in contrast to the analyses of 
Refs.~\cite{beneke-Bdpi0K0,SCET-Bdpi0K0}.

\begin{figure}
\begin{center}
\includegraphics[width=12cm]{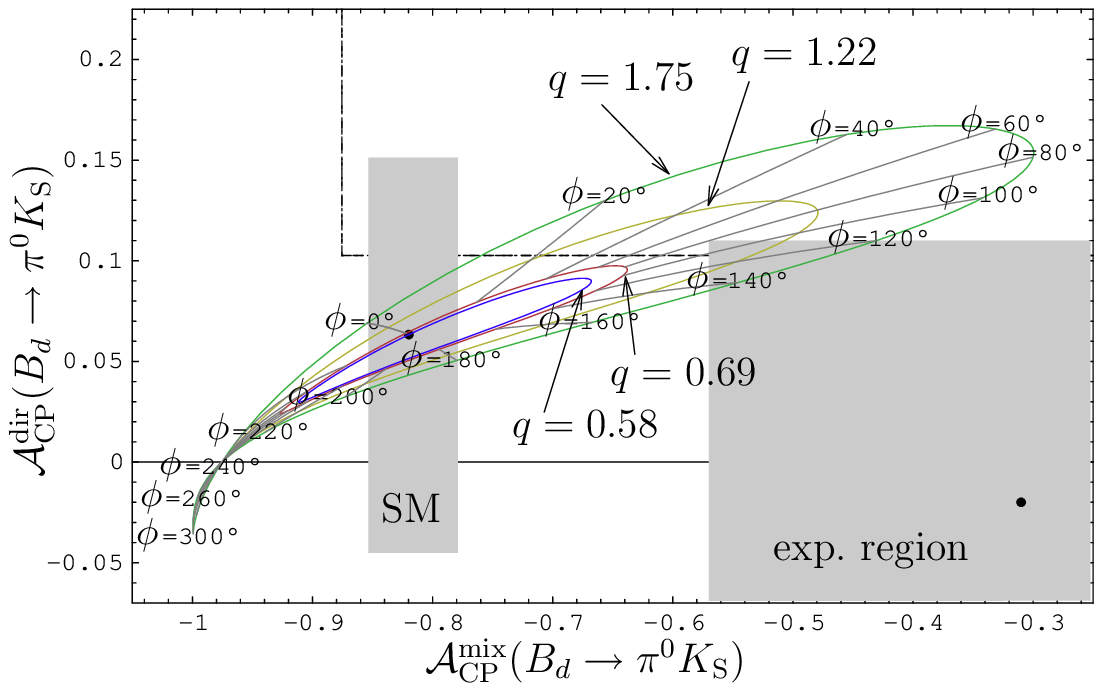}
\end{center}
\caption{The situation in the 
${\cal A}_{\rm CP}^{\rm mix}(B_d\to\pi^0 
K_{\rm S})$--${\cal A}_{\rm CP}^{\rm dir}(B_d\to\pi^0 K_{\rm S})$ plane:
we show contours for values of $q=0.58$ to $q=1.75$ with $\phi \in
[0^\circ,360^\circ]$. The grey area represents the 1$\sigma$ experimental
range (see (\ref{Bdpi0K0-exp})), and the central value is indicated 
by the black dot.
\label{fig:Adirpi0KS-Amixpi0KS}}
\end{figure}

Let us now turn to our NP scenario. Using the modified parameters of 
$(q,\phi)$ in (\ref{q-phi}) yields the following results:
\begin{equation}
{\cal A}_{\rm CP}^{\rm dir}(B_d\to\pi^0K_{\rm S}) 
   = 0.01\,^{+0.14}_{-0.18}, 
   \quad
{\cal A}_{\rm CP}^{\rm mix}(B_d\to\pi^0K_{\rm S}) 
   = -(0.96\,^{+0.04}_{-0.08}).
\end{equation}
Consequently, as already noticed in \cite{BFRS}, these specific EW penguin
parameters imply an enhancement of $\Delta S$ with respect to the SM case: 
\begin{equation}\label{DSBFRS}
\Delta S= 0.27\,^{+0.05}_{-0.09}.
\end{equation}
Thus the best values for $(q,\phi)$ that are required to confront the 
small value of $R_{\rm n}$ with the theoretical interpretation within
our strategy make the disagreement with the data for
${\cal A}_{\rm CP}^{\rm mix}(B_d\!\to\!\pi^0 K_{\rm S})$ even larger 
than in the SM. The question then arises whether there exist values of
$(q,\phi)$ for which $\Delta S$ could be smaller than in the SM or even 
reverse the sign. As seen already in Fig.~10 of \cite{BFRS} and in its 
updated version in Fig.~\ref{fig:Adirpi0KS-Amixpi0KS} here, such values of 
$(q,\phi)$ can indeed be found. We will return to this issue after the
constraints from rare decays have been taken into account. In view of 
the large experimental errors of the mixing-induced CP asymmetry
of the $B_d\to\pi^0K_{\rm S}$ channel, it is unfortunately not possible to
draw definite conclusions at the moment. 

\begin{figure}
\begin{center}
\includegraphics[width=12cm]{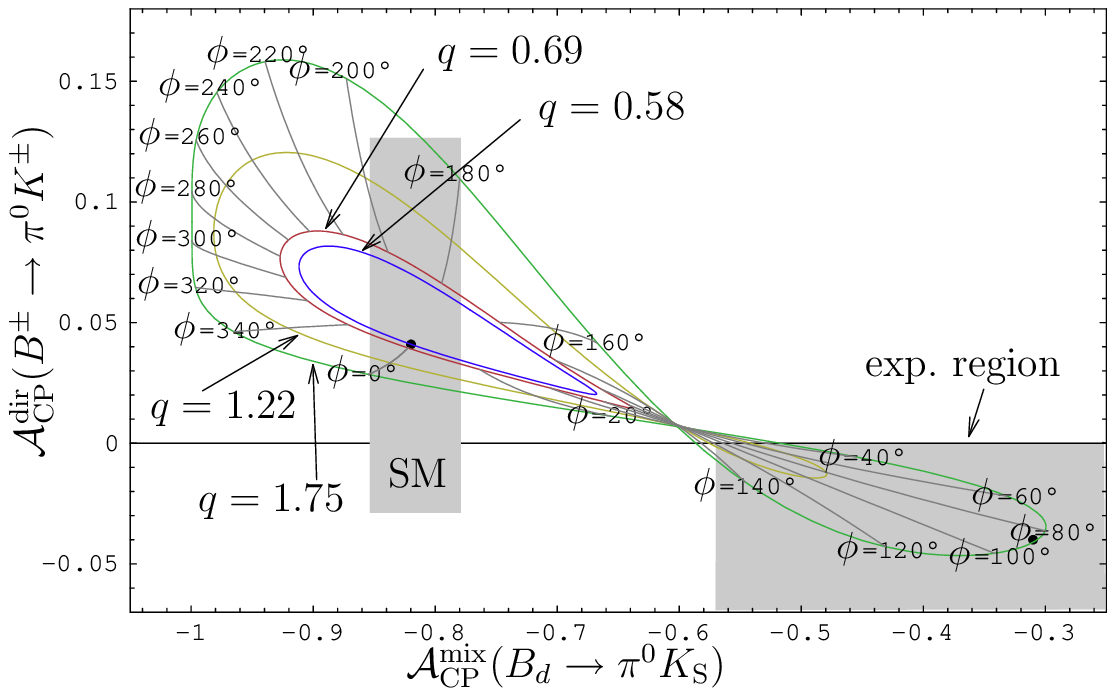}
\end{center}
\caption{The situation in the 
${\cal A}_{\rm CP}^{\rm mix}(B_d\to\pi^0K_{\rm S})$--${\cal A}_{\rm CP}^{\rm dir}
(B^\pm\to\pi^0K^\pm)$ plane, in analogy to Fig.~\ref{fig:Adirpi0KS-Amixpi0KS}.
\label{fig:Adirpi0KS-Amixpi0K+}}
\end{figure}

Finally, there is still one CP asymmetry of the  $B\to\pi K$ system left:
\begin{equation}\label{AdirBppi0KP-exp}
{\cal A}_{\rm CP}^{\rm dir}(B^\pm\to\pi^0K^\pm) \,\stackrel{{\rm exp}}{=} 
-0.04\pm 0.04.
\end{equation}
This quantity received also a lot of attention, in particular as its experimental 
value differs from 
\begin{equation}\label{AdirBdpimKp-exp}
{\cal A}_{\rm CP}^{\rm dir}(B_d\to\pi^\mp K^\pm) \,\stackrel{{\rm exp}}{=} 
0.115\pm 0.018,
\end{equation}
which we have used in (\ref{H-rel}). 
On the other hand, both asymmetries are 
expected to be equal in the naive limit of vanishing colour-suppressed 
tree and electroweak penguin topologies. The lifted colour-suppression 
shown through the large value of $x$ could, in principle, be responsible 
for this difference, but, calculating this asymmetry in the SM and our NP 
scenario (\ref{q-phi}), we find:
\begin{equation}\label{AdirBppi0KP}
{\cal A}_{\rm CP}^{\rm dir}(B^\pm\to\pi^0K^\pm)|_{\rm SM} 
= 0.04\,^{+0.09}_{-0.07},
\quad
{\cal A}_{\rm CP}^{\rm dir}(B^\pm\to\pi^0K^\pm)|_{\rm NP}
= 0.09\,^{+0.20}_{-0.16},
\end{equation}
so that the SM still prefers a positive value of ${\cal A}_{\rm CP}^{\rm dir}(B^\pm\to\pi^0K^\pm)$.
In view of the large uncertainties, no stringent test is provided at this point.
Nevertheless, it is tempting to play a bit with this asymmetry. In analogy to
Fig.~\ref{fig:Adirpi0KS-Amixpi0KS}, we show in Fig.~\ref{fig:Adirpi0KS-Amixpi0K+}
the situation in the 
${\cal A}_{\rm CP}^{\rm mix}(B_d\to\pi^0K_{\rm S})$--${\cal A}_{\rm CP}^{\rm dir}
(B^\pm\to\pi^0K^\pm)$ plane for various values of $q$ with $\phi\in[0^\circ,360^\circ]$.
We observe that also the current experimental value of the CP asymmetry of the 
charged $B^\pm\to\pi^0K^\pm$ mode seems to show a preference for 
positive values of $\phi$ around $+90^\circ$. It will be interesting to monitor
these topics as the data improve. We will return to this issue in 
Section \ref{sec:rare-BK-scenarios}.


\boldmath
\subsection{A Closer Look at $SU(3)$-Breaking Effects}\label{ssec:SU3break}
\unboldmath
Before leaving the $B\to\pi K$ system, let us have a critical look at the sensitivity
of our results on large (non-factorizable) corrections to the working assumptions
listed in Section~\ref{sec:intro}. As we discussed in detail in \cite{BFRS,BFRS-up},
internal consistency checks of these assumptions are provided within our
strategy. An example is relation (\ref{H-rel}). These checks are
currently satisfied at the $25\%$ level, and can be improved systematically with
better experimental data. Consequently, no violation of our working
assumptions is indicated. On the other hand, since sizeable non-factorizable
$SU(3)$-breaking effects cannot yet be excluded, let us investigate their
impact on our numerical results.

In our analysis of the $B\to\pi\pi,\pi K$ system, we include
factorizable $SU(3)$-breaking corrections through appropriate form-factor
and decay-constant ratios. The relevant relation is Eq.~(3.55) of Ref.~\cite{BFRS-up},
which relates the parameters $(x,\Delta)$ of the $B\to\pi\pi$ system to their
$B\to\pi K$ counterparts:
\begin{equation}\label{x-rel-SU3}
x'e^{i\Delta'}=\left[\frac{f_\pi F_{BK}(M_\pi^2;0^+)}{f_K F_{B\pi}
(M_K^2;0^+)}\right]xe^{i\Delta} 
\equiv \rho_{SU(3)} \, xe^{i\Delta} \,.
\end{equation}
From light-cone sum-rules \cite{Balletal}, it was found that 
$\rho_{SU(3)}^{\rm fact}=1.05\pm0.18$. This factor is also included 
in the updated analysis presented in this paper. In order to explore the 
impact of large non-factorizable $SU(3)$-breaking effects on our analysis,
we will use $|\rho_{SU(3)}|= 1.05\pm0.36$, i.e.\ enlarge the error of 
$|\rho_{SU(3)}^{\rm fact}|$ by 100\%, and also allow for a CP-conserving 
strong phase of $\rho_{SU(3)}$ between $-15^\circ$ and $+15^\circ$. 
Concerning the relation of the $B^0_d\to\pi^+\pi^-$ parameters $(d,\theta)$
to their $B^0_d\to\pi^-K^+$ counterparts $(d',\theta')$, we follow 
\cite{gamma-det,FleischerMatias}, and introduce $SU(3)$-breaking parameters
through
\begin{equation}
d'=\xi d, \quad \theta'=\theta+\Delta\theta.
\end{equation}
In the numerical analysis, we  consider then $\xi= 1.0\pm0.18$,
and allow the strong phase $\Delta\theta$ to vary freely between 
$-15^\circ$ and $+15^\circ$. 

The impact of this conservative treatment of non-factorizable $SU(3)$-breaking
corrections on our SM analysis of the $B\to\pi K$ system is surprisingly small, as 
can be seen in Table~\ref{tab:SU3}. Even with significantly enhanced uncertainties, 
it is not possible to accommodate the whole $B\to\pi K$ data in a satisfactory manner
within the SM, and the $B\to\pi K$ puzzle persists. Consequently, it is in fact
very exciting to speculate on NP effects, as we have done in 
Subsections~\ref{ssec:BpiK-EWP} and \ref{ssec:BpiK-CP}. 
Let us next explore the interplay with rare $K$ and $B$ decays.

\begin{table}
\vspace{0.4cm}
\begin{center}
\begin{tabular}{|c||c|c|}
\hline
  Quantity &  Default values & Non-fact.\ $SU(3)$ breaking  \\ \hline 
  $\gamma$ & $(73.9^{+5.8}_{-6.5})^\circ$ & $(73.9^{+9.4}_{-9.0})^\circ$ \\ \hline 
  $R$         & $ 0.96\pm0.02 $ & $0.96\,^{+0.03}_{-0.04} $  \\ \hline 
  $R_{\rm c}$ & $ 1.15\pm0.05 $ & $1.15\pm0.07$ \\ \hline
  $R_{\rm n}$ & $ 1.12\pm0.05 $ & $1.12\pm0.06$ \\ \hline
  ${\cal A}_{\rm CP}^{\rm dir}(B^\pm\to\pi^0K^\pm)$ 
   & $0.04\,^{+0.08}_{-0.07}$ & $0.04\,^{+0.13}_{-0.11}$ \\ \hline
  ${\cal A}_{\rm CP}^{\rm dir}(B_d\to\pi^0K_{\rm S})$ 
   & $0.06\,^{+0.09}_{-0.10}$ & $0.06\,^{+0.13}_{-0.16}$ \\ \hline
  $\Delta S$ & $0.13\pm0.05$ & $0.13\,^{+0.06}_{-0.07}$ \\ \hline
\end{tabular}
\end{center}
\caption{The impact of large non-factorizable $SU(3)$-breaking effects
on our SM analysis. The ``default'' results of our analysis include
factorizable $SU(3)$-breaking corrections, as described in the text.}\label{tab:SU3}
\end{table}

\boldmath
\section{Interplay with Rare $B$ and $K$ Decays and Possible Future 
Scenarios}\label{sec:rare-BK-scenarios}
\unboldmath
An attractive feature of the approach in \cite{BFRS,BFRS-up} is a 
direct connection between non-leptonic $B$ decays and rare $B$ and $K$ 
decays \cite{BFRS-I}. Assuming that the dominant NP contributions enter 
through the $Z^0$-penguin function $C$, and using the renormalization-group 
evolution from scales $\ord(M_W,m_t)$ to scales $\ord(m_b)$, we can directly 
investigate the impact of the modified EW penguin contributions in the
$B\to\pi K$ modes on rare $B$ and $K$ decays.

Proceeding in this manner we find that the value of $(q,\phi)$ in 
(\ref{q-phi}), which is preferred by the $B\to\pi K$ observables $R_{\rm n,c}$, 
requires the one-loop short-distance functions $X$ and $Y$ to be at least as 
high as
\begin{equation}\label{XY1}
|X|_{\rm min}\approx 
|Y|_{\rm min}\approx 2.2,
\end{equation}
to be compared with $X\approx 1.5$ and $Y\approx 1.0$ in the SM.

The values in (\ref{XY1}) appear to violate the $95\%$ probability 
upper bounds
\begin{equation}\label{XY2}
X\le 1.95, \qquad Y\le 1.43,
\end{equation}
obtained recently in the context of minimal flavour violation (MFV) 
\cite{Bobeth:2005ck}. While
our scenario of NP having new complex phases goes beyond MFV, the 
inspection of the known formulae for $B\to X_s l^+l^-$ shows that the upper 
bound on $Y$ in (\ref{XY2}) is difficult to avoid if the only NP
contribution resides in the EW penguins and the operator basis is the same as
in the SM. For our analysis below we will, therefore, use an only slightly 
more generous bound and impose $\left|Y \right| \leq 1.5$. Taking then those 
values of $(q,\phi)$ from (\ref{q-phi}) that also satisfy $\left|Y \right|=1.5$ leaves 
us with
\begin{equation}
\label{q-phi-RD}
q= 0.48 \pm 0.07 ,\quad \phi=-(93 \pm 17 )^\circ.
\end{equation}
Note that this corresponds to a modest {\it suppression} of the 
magnitude of the EW penguin parameter relative to its new SM value of 0.58.

Another possible solution to the clash between (\ref{XY1}) and (\ref{XY2}) would be
the introduction of new complex phases in the photon magnetic penguin 
contribution that has no impact on the $B\to \pi K$ decays but can influence 
$B\to X_s l^+l^-$. This could weaken the tension between
(\ref{XY1}) and (\ref{XY2}) subject to the bounds on the  CP asymmetry in the 
$B\to X_s\gamma$ decay, where the photon magnetic penguin plays an
important r\^ole \cite{Kagan:1998bh}. 
Another avenue one could explore would be the introduction of
new operators in $B\to X_s l^+l^-$ that would invalidate the bounds in 
(\ref{XY2}). For instance, new operators originating in Higgs penguins 
in the MSSM with a large $\tan\beta$ could help here. The impact of these
new operators on $B\to X_s l^+l^-$ turns out to be moderate when the
constraints on their Wilson coefficients from $B_s\to\mu^+\mu^-$ are taken 
into account \cite{Hiller:2003js}. Still their presence can definitely weaken the 
bounds in (\ref{XY2}), so that the values in (\ref{XY1}) are compatible with rare 
decay constraints in such a more complicated NP scenario.

In spite of these possibilities, we will not explore them in the present 
paper because the predictive power of this more general NP scenario is 
significantly smaller than of our scenario, unless a specific model
is considered.
Instead we will investigate how various modifications of $(q,\phi)$, which
allow us to satisfy the bounds in (\ref{XY2}), influence our results for 
the observables of the $B\to \pi K$ system presented in Section \ref{sec:BpiK},
and the predictions for rare decays discussed in detail in \cite{BFRS,BFRS-up}.
For this purpose, we have introduced three scenarios that represent possible 
future measurements of $R_{\rm n}$ and $R_{\rm c}$:
\begin{itemize}
\item Scenario A: $q=0.48$, $\phi = -93^{\circ}$, which is compatible with 
the present $B \to \pi K$ data and the rare decay bounds 
(see (\ref{q-phi-RD})).
\item Scenario B: we assume that $R_{\rm n}$ goes up, and take $q=0.66,$
$\phi=-50^{\circ}$, which leads to 
$R_{\rm n} = 1.03 $, $R_{\rm c}=1.13$ and some interesting effects 
in rare decays, as we shall see below. This would, for example, occur if radiative
corrections to the  $B_d^0\to\pi^- K^+$ branching ratio enhance 
$R_{\rm n}$ \cite{Baracchini:2005wp}, though this alone would probably 
account for only about $5\%$.
\item Scenario C: assume that both $R_{\rm n}$ and $R_{\rm c}$ move towards 1;
taking $R_{\rm n}=R_{\rm c}=1$ leads to $q=0.54$, 
$\phi=61^{\circ}$. The {\it positive} sign of the phase in this scenario distinguishes it strongly from both others.
\end{itemize}

The result of this exercise is contained in Tables \ref{Scentab1} and \ref{Scentab2}:
in Table \ref{Scentab1}, we show the values of a number of observables of the 
$B\to\pi K$ system in the three scenarios, while in Table~\ref{Scentab2}, we show 
the corresponding values of the most interesting branching ratios for rare $K$ and 
$B$ decays. To this end, we have used for the angle $\beta$ the value of 
$\beta_{\rm true}$ in Table~\ref{tab:RUT}.
We observe that, in particular, the interplay of the $K \to \pi \bar \nu \nu$ 
modes is a very good and clean indication of which
kind of NP scenario to look for. Due to the interference of charm and top 
contributions in $K^+ \to \pi^+ \bar \nu \nu$, it is also 
the decay that can most naturally be suppressed (though this is in contrast to 
the present experimental value). On the other hand, 
$\mbox{BR}(K_{\rm L} \to \pi^0 \bar \nu \nu)$ is always enhanced due to the large 
values  of $\phi$ and the absence of the charm contribution.
Concerning the observables of the $B \to \pi K$ system,
${\cal A}_{\rm CP}^{\rm mix}(B_d\!\to\!\pi^0 K_{\rm S})$ offers a particularly
interesting probe. This CP asymmetry comes out very large in Scenarios 
A and B, where the NP phase is negative. On the other hand, the positive sign 
in Scenario C brings this value closer to the data, in accordance with
the features pointed out in Subsection~\ref{ssec:BpiK-CP}.
Similarly the experimental value of
\begin{equation}
\Delta A \equiv {\cal A}_{\rm CP}^{\rm dir}(B^\pm\to\pi^0K^\pm)
               -{\cal A}_{\rm CP}^{\rm dir}(B_d\to\pi^\mp K^\pm)
\,\stackrel{{\rm exp}}{=}  -0.16\pm0.04  \label{DeltaA}
\end{equation}
favours a positive value of $\phi$.

\begin{table}[hbt]
\vspace{0.4cm}
\begin{center}
\begin{tabular}{|c||c|c|c|c|c|}
\hline
  Quantity & SM & Scen A & Scen B &  Scen C & Experiment
 \\ \hline
$R_{\rm n}$  & 1.12 &$0.88$ & 1.03 &  1 & $0.83 \pm 0.08$ \\\hline
$R_{\rm c}$  & 1.15 &$0.96$ & 1.13 & 1  & $1.01 \pm 0.09$ \\\hline
${\cal A}_{\rm CP}^{\rm dir}(B^\pm\!\to\!\pi^0 K^\pm) $ &
  0.04 & $0.07$  \rule{0em}{1.05em}& 0.06 & 0.02  &  $-0.04 \pm 0.04$ \\ \hline
${\cal A}_{\rm CP}^{\rm dir}(B_d\!\to\!\pi^0 K_{\rm S})$ & 
  0.06 & $0.04$  \rule{0em}{1.05em}& 0.03  & 0.09 & $-0.02 \pm 0.13$ \\ \hline 
${\cal A}_{\rm CP}^{\rm mix}(B_d\!\to\!\pi^0 K_{\rm S})$ & 
  $-0.82$ & $-0.89$\rule{0em}{1.05em}& $-0.91$ & $-0.70$ &  $-0.31 \pm 0.26$ \\ \hline
$\Delta S$ & 0.13& 0.21& 0.22& 0.01& $-0.38\pm0.26$ \\ \hline
$\Delta A$ & $-0.07$& $-0.04$& $-0.05$& $-0.09$& $-0.16\pm0.04$ \\ \hline
\end{tabular}
\caption{\label{Scentab1} The $B\to\pi K$ observables for the 
 three scenarios introduced in the text. }
\end{center}
\end{table}

\begin{table}[hbt]
\begin{center}
\begin{tabular}{|c||c|c|c|c|c|}
\hline
  Decay & \quad SM \quad &   Scen A &   Scen B &   Scen C &
  \parbox{2.3cm}{\rule{0em}{1em}Exp. bound \\90\% {\rm C.L.}}
 \\ \hline
$\mbox{BR}(K^+ \to \pi^+ \bar \nu \nu)/10^{-11}$  &   
 $ 9.3$ & $2.7 $ &  $8.3 $ & $8.4 $ &  $(14.7^{+13.0}_{-8.9}) $\rule{0em}{1.05em} \\ \hline
$\mbox{BR}(K_{\rm L} \to \pi^0 \bar \nu \nu)/10^{-11}$  &
 $ 4.4$ &  $ 11.6$ &  $27.9$ & $7.2$ &  $ < 2.9 \cdot10^{4} $ \\ \hline
$\mbox{BR}(K_{\rm L} \to \pi^0  e^+ e^-)/10^{-11}$ &  
 $ 3.6$ & $4.6$  &    $7.1$ &  $4.9 $&   $<28$ \\ \hline
$\mbox{BR}(B \to X_s\bar \nu \nu)/10^{-5}$  &  
 $3.6$  &  $ 2.8 $&   $4.8$ &  $3.3 $ &  $<64$ \\ \hline
$\mbox{BR}(B_s \to \mu^+ \mu^-)/10^{-9}$  & 
 $3.9$ & $9.2$ & $ 9.1$ &  $7.0 $&  $<1.5\cdot 10^{2}$ \rule{0em}{1.05em}\\ \hline 
$\mbox{BR}(K_{\rm L} \to \mu^+ \mu^-)_{\rm SD}/10^{-9}$ &  
 $ 0.9$ & $0.9$  &    $0.001$ &  $0.6 $&   $<2.5$ \\ 
\hline
\end{tabular}
\caption{\label{Scentab2} Rare decay branching ratios for the three scenarios 
introduced in the text.}
\end{center}
\end{table}

In Table~\ref{Scentab2}, we show the effect of the various scenarios
on selected rare decays. Our SM results differ slightly from the
standard values because we use the CKM input from
the first column in Table~\ref{tab:RUT}. Larger values of $R_t$ and $\bar\eta$ 
than found in \cite{UTfit,CKMfitter} result in higher values of 
$\mbox{BR}(K^+ \to \pi^+ \bar \nu \nu)$
and $\mbox{BR}(K_{\rm L} \to \pi^0 \bar \nu \nu)$, respectively, than found
in \cite{BGHN} and \cite{BSU}. The great sensitivity of the rare decay
branching ratios to the parameters $(q,\phi)$ demonstrates clearly 
the impressive power of rare $K$ and $B$ decays to search for NP effects in
the EW penguin sector. 

\boldmath
\section{Conclusions}\label{sec:concl}
\unboldmath
In this paper, we have reconsidered our analysis of $B\to\pi\pi,\pi K$
and rare $B$ and $K$ decays in view of the new $B$-factory data 
for $(\sin2\beta)_{\psi K_{\rm S}}$ and the two-body modes as well as more 
stringent bounds on rare decays. The main new messages from our analysis 
are as follows:
\begin{itemize}
\item The $B_d\to\pi^+\pi^-$ and
$B_d\to\pi^\mp K^\pm$ modes, which are marginally affected by EW penguins, 
allow us to determine $\gamma=(74\pm6)^\circ$. Complementing 
this value with $|V_{ub}/V_{cb}|$, we can determine the true unitarity triangle, 
allowing us to search for NP contributions to $B^0_d$--$\bar B^0_d$ mixing; we 
find a NP phase $\phi_d^{\rm NP}=-(8.2\pm3.5)^\circ$. 

\item  The $B\to\pi K$ puzzle, which is in particular reflected by the low 
experimental value of the ratio $R_{\rm n}$ of the neutral $B\to\pi K$ branching 
ratios, persists. It still points to NP in the EW penguin sector, and favours
a large NP phase $\phi\sim -90^\circ$, although now also a value around
$+90^\circ$ can bring us rather close to the current experimental ranges
of $R_{\rm n,c}$. 

\item $\phi\sim +90^\circ$ would allow us to accommodate also the 
pattern of $(\sin 2\beta)_{\pi^0 K_{\rm S}}<(\sin2\beta)_{\psi K_{\rm S}}$, which 
may be indicated by the $B$-factory data and received recently a lot of attention,
although the measurements suffer still from large uncertainties. A similar comment 
applies to the difference $\Delta A$ in (\ref{DeltaA})
of the direct CP asymmetries of the $B_d\to\pi^\mp K^\pm$ 
and $B^\pm\to\pi^0K^\pm$ modes, which could be increased 
for $\phi\sim +90^\circ$. However, the latter CP asymmetry suffers from large uncertainties 
in our approach and does therefore not (yet) allow a stringent test.

\item The internal consistency checks of the working assumptions of our
strategy are satisfied at the level of $25\%$, and can be improved through
better data. We studied the sensitivity of our numerical predictions of the $B\to\pi K$
observables on non-factorizable $SU(3)$-breaking effects of this order 
of magnitude, and found that the impact is surprisingly small. Consequently,
it is in fact very exciting to speculate on NP effects in the EW penguin contributions
to the $B\to\pi K$ decays. 
\item
In view of the fact that the parameters $(q,\phi)$ needed for the explanation
of the low value of $R_{\rm n}$ appear to imply rare decay branching ratios 
that violate the experimental bound from $B\to X_s l^+l^-$, we have explored 
various scenarios for $(q,\phi)$ that allow us to satisfy the rare decay
constraints, but still give interesting results for both the $B\to \pi K$ decays 
and the rare $K$ and $B$ decays. Needless to say, our analysis of the $B\to\pi\pi$ 
modes and the determination of the angle $\gamma$ described above are not 
affected by these modifications. On the other hand, we find again that reversing 
the sign of the NP phase $\phi$ brings  the mixing-induced asymmetry 
of $B_d\to \pi^0 K_{\rm S}$ closer to the data.
\end{itemize}

 Our analysis demonstrates that the simultaneous study of non-leptonic 
 $B$-decay branching ratios, the corresponding direct and mixing-induced CP
 asymmetries and rare $K$ and $B$ decays within a consistent phenomenological 
 framework developed in \cite{BFRS}, can with improved data shed light on new
 physics and the structure of QCD dynamics in non-leptonic $B$ decays.

\vspace*{0.5truecm}

\noindent
{\bf Acknowledgements}\\
\noindent
This work has been supported in part by the
Bundesministerium f\"ur
Bildung und Forschung under the contract 05HT4WOA/3 and by the German-Israeli
Foundation under the contract G-698-22.7/2002.

\end{document}